# A New Android Malware Detection Approach Using Bayesian Classification

Suleiman Y. Yerima, Sakir Sezer,
Gavin McWilliams
Centre for Secure Information Technologies
Queen's University Belfast,
Belfast, Northern Ireland, United Kingdom
E-mail: {s.yerima, s.sezer, g.mcwilliams}@qub.ac.uk

Igor Muttik
Senior Research Architect
McAfee Labs
London, United Kingdom
E-mail: igor_muttik@mcafee.com

*Abstract*— **Mobile malware has been growing in scale and complexity as smartphone usage continues to rise. Android has surpassed other mobile platforms as the most popular whilst also witnessing a dramatic increase in malware targeting the platform. A worrying trend that is emerging is the increasing sophistication of Android malware to evade detection by traditional signature-based scanners. As such, Android app marketplaces remain at risk of hosting malicious apps that could evade detection before being downloaded by unsuspecting users. Hence, in this paper we present an effective approach to alleviate this problem based on Bayesian classification models obtained from static code analysis. The models are built from a collection of code and app characteristics that provide indicators of potential malicious activities. The models are evaluated with real malware samples in the wild and results of experiments are presented to demonstrate the effectiveness of the proposed approach.**

## I. INTRODUCTION

The Android mobile platform is increasing in popularity surpassing rivals like iOS, Blackberry, Symbian and Windows mobile. It is estimated that there are currently around 675,000 applications in the official Google's Android market, with an estimated 25 billion downloads (as at October 2012) [1]. At the same time, malware targeting the Android platform has risen sharply over the last two years. According to a report from Fortinet (issued in November 2011), there exist approximately 2000 Android malware samples belonging to 80 different families [2].

Since discovery of the first Android malware in August 2010 [3], new families have evolved in sophistication and are becoming increasingly difficult to detect by traditional signature-based anti-virus. More recent families have been observed to exhibit polymorphic behavior, increased code obfuscation, encryption of malicious payloads, as well as stealthy command and control communication with remote servers. In fact, some Android malware like AnserverBot are known to have the capability to fetch and execute payload at run time thus rendering its detection quite challenging.

Security experts believe that difficulties in spotting malicious mobile applications results in most Android malware remaining unnoticed for up to 3 months before discovery [2]. Furthermore, Oberheide et al. [4] observed that it took on average 48 days for a signature-based antivirus engine to become capable of detecting new threats.

Clearly, there is a need for more effective detection solutions to overcome the aforementioned challenges and mitigate the impact of evolving Android malware. Hence, in this paper we present a new approach - based on Bayesian classification - which utilizes certain characteristic features frequently observed with malware samples to classify an app as 'suspicious' or 'benign'. Our approach is developed as a proactive method aimed at uncovering known families as well as unknown malware so as to reduce incidents of malware in marketplaces from evading detection.

The Bayesian classification based approach can complement signature-based scanning thus enabling harmless apps obtained from Android marketplaces to be verified whilst isolating suspicious samples for further scrutiny. Thus, it is a viable tool for filtering the vast amount of apps added online on a daily basis; estimated to average more than 1200 a day [5], while narrowing the window of opportunity to reach end user devices and wreak havoc. The approach is not only useful for prioritizing apps that may require further scrutiny but is also a very effective tool for uncovering unknown malware.

The rest of the paper is organized as follows: related work is discussed followed by background on the Android system. Next, the Android reverse engineering and static analyses that underpins our classification approach is presented followed by the Bayesian classification model formulation. Experiments, results and analyses follow; we then conclude the paper and outline future work.

## II. RELATED WORK

Different from earlier work on mobile malware detection that rely on on-device anomaly [6] or behavioral based detection [7], [8], our approach is off-device, and employs static analysis of Android application packages. Hence it is aimed at detecting and filtering potentially malicious apps before they installed or run on a device. Additionally, our approach aims to close the window of opportunity for undetected malware in marketplaces, whilst also avoiding the challenges of device performance bottlenecks.

Static analysis has the advantage of being undetectable, as obviously malware cannot modify its behavior during analysis [2]. Thus, it has been applied to vulnerability assessment, profiling and malware detection for the Android platform. For example, Comdroid [9] is a static analysis tool for application communication based vulnerabilities in Android. Similarly, DroidChecker [10] is an Android application analysis tool which searches for capability leakage vulnerability in Android applications. ProfileDroid [11], RiskRanker [5], and the tool presented in [12], all leverage static analysis for profiling and analyzing Android applications. Other existing works that employ static analysis for detection of malicious activities like SCANDAL [13], AndroidLeaks [14], and the framework presented in [15], focus on privacy information leakage. Our work on the other hand, covers detection of a wider scope of malicious activities than privacy information loss.

In [16] Blasing et al. presented an Android Application Sandbox (AAS) that uses both static and dynamic analyses on Android applications to automatically detect suspicious applications. Compared to AAS, our approach covers a much wider range of pattern attributes extracted not only from the application code logic but also scrutiny of resources, assets, and executable libraries where malicious payload could be lurking. Additionally, these attributes contribute to a ranked feature set which drives our Bayesian classification model.

In [2] Apvrille and Strazzere employ a heuristics approach based on static analysis for Android malware detection. Their heuristic engine uses 39 different flags and then outputs a risk score to highlight the most likely malicious sample. Our approach shares similarity in the reverse engineering technique, but differs by utilizing a machine learning based method that offers more flexibility. In [17], Schmidt et al. employ static analysis on executables to extract their function calls using the *readelf* command. They then compare these function call lists with those from Linux malware executable in order to classifying the executables using learning algorithms. In contrast, our static analysis approach is based on automated analyses of Android packages. Moreover, Android malware samples across a range of existing families are employed in our work rather than Linux malware executables.

Other earlier non-Android based papers have explored data mining and machine learning techniques for malware identification including for example [18] and [19]. For the Android platform, a recent paper by Sahs and Khan [20] presented a machine learning approach to Android Malware detection based on SVM. A single-class SVM model derived from benign samples alone is used. The approach in our paper differs in that a more extensive set of code-based and external attributes provide features for training the model whereas [20] used the Android permissions in the Manifest files. Furthermore, unlike in [20], our classification models are trained with both benign samples and a range of samples from across 49 malware families discovered in the wild.

In summary, the main contributions of this paper different from existing related works in the literature are as follows:
- A novel application of a machine learning technique, i.e. Bayesian classification, for signature-free Android malware detection using static code analysis.
- Model building and in-depth empirical evaluation with a range of representative malware samples from across 49 existing malware families in the wild.

Our work is motivated by the sheer amount of Android applications that require scanning and the rate at which new ones keep appearing in different app markets. We also note that there is significant delay between malware release and its detection which even suggests that we are currently unaware of several Android malware in the wild [2].

### III. ANDROID SYSTEM ARCHITECTURE

Android is designed for mobile devices with resource constraints. It provides a sandboxed application execution environment where a customized embedded Linux system interacts with the phone hardware and off-processor cellular radio while the Binder middleware and application API runs on top of Linux. As shown in Figure 1, Android is effectively a software stack for mobile devices that includes an operating system, middleware and key applications and uses a modified version of the Linux kernel. The Linux 2.6.x kernel lying at the foundation of the Android platform serves as a hardware abstraction layer, which offers an existing memory management, process management, security and networking model upon which the rest of the Android platform is built. The native libraries such as SQLite, Webkit and SSL, layered on top of the Linux kernel, provide most of the functionality of the Android system. The Application framework layer provides all the APIs that the applications require to access the device hardware: location information, running background services, etc. The application's only interface to the phone is through these API's. Each application is executed within a Dalvik Virtual Machine (DVM) running under a unique UNIX uid.

At the higher operating system layers we have the user applications such as the phone application, home application, etc. which come pre-installed and other applications that are downloaded from the Google Play market or alternative marketplaces or even installed manually from .apk files. These additional apps extend the functionality of the smartphone and pose potential threat to user security and privacy if they happen to be malicious.

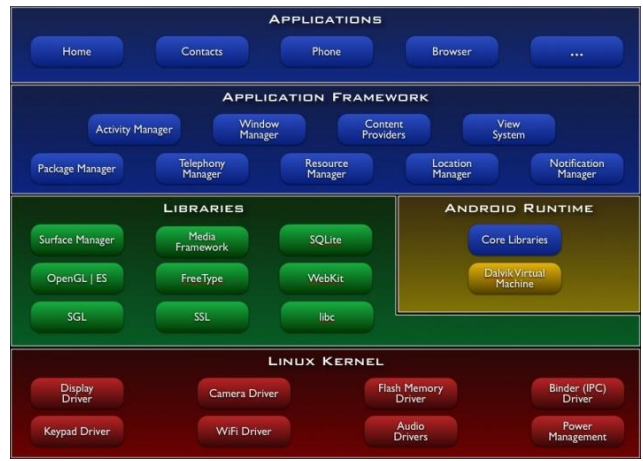

Figure 1. Android System Architecture [21].

*A. Android application basics*

Android applications (or apps) are written in the Java programming language. The Android SDK tools compile the code—along with any data and resource files—into an Android package, an archive file with an .apk suffix. All the code in a single .apk file is considered to be one application and it is this file that Android-powered devices use to install the app.

The Android app is built from four different types of components: Activities, Services, Broadcast Receivers, and Content Providers [22]. An app must declare its components in a manifest file which must be at the root of the application project directory. Before the Android system can start an application component, the system must know that the component exists by reading the application's manifest file. The manifest file also states the user permissions that the application requires, such as internet access or read-access to the user's contacts.

Android apps are distributed as self-contained packages called APKs. An APK (Android Package) is a compressed (ZIP) bundle of files typically consisting of: AndroidManifest.xml (manifest file), classes.dex (A single file which holds the complete bytecode to be interpreted by Dalvik VM). Other binary or XML-based resources required by the app to run may be held in res/ and assets/ folders.

The detection strategy developed in this paper leverages the applications' reliance on the platform APIs and their structured packaging to extract certain properties that could serve as indicators of suspicious activity, such as intent to exfiltrate sensitive information, launch a malicious payload at runtime, or presence of embedded secondary payload in the external folders etc. These properties then form the basis of our Bayesian classifier, which is used to determine whether a given Android app is harmless or suspicious.

*B. Application Reverse Engineering*

In order to obtain the feature sets for building the Bayesian model, we implemented a Java-based Android package profiling tool for automated reverse engineering of the APK files. The steps involved are shown in Figure 2.

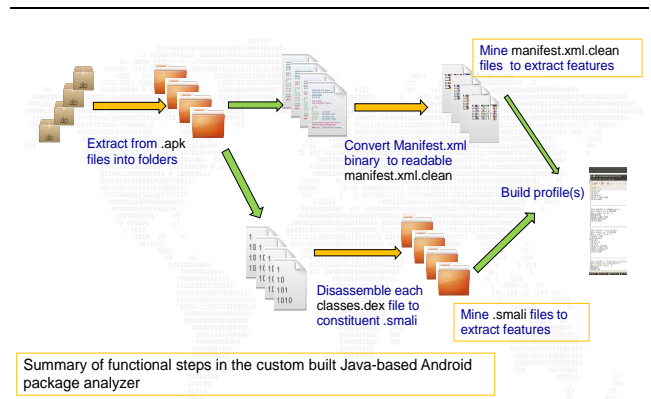

Figure 2. Automated reverse engineering of the Android apps with the Java-based APK analyzer.

To parse the .dex file, a tool called Baksmali [23] is used, which is a disassembler for the dex format used by Dalvik. Baksmali disassembles .dex files into multiple files with .smali extensions. Each .smali file contains only one class information which is equivalent to a Java .class file.

*C. Applying property detectors to reverse engineered packages*

After reverse engineering an APK, a set of detectors are applied within the Java-based APK analyzer to check for properties which are then mapped into feature vectors for the Bayesian classifier. The properties matched by the detectors include API calls, Linux system commands and permissions contained in the manifest files. Other properties such as encryption of code, and the presence of secondary .apk or .jar files are also matched.

*1) API call detectors*

Considering API calls provide the means for applications to interact with the phone, their observation via static code inspection can provide insight into the application's intended runtime activities. This allows the feature extractor to build a profile of the apps based on the API calls which when used in conjunction with other properties facilitates the discovery of malicious intent via the Bayesian Classifier's operation.

For example API calls to the Android Telephony manager which is used to retrieve subscriber ID, phone ID, and other similar information are monitored. Similarly, API calls within the code, such as those for receiving/sending SMS, calling phone numbers, retrieving telephony information, listing or installing other packages on the phone, etc. are further properties which provide additional features for the classifier.

*2) Command detectors*

API call detectors scrutinize the reverse engineered .dex files to extract patterns for feature selection. In order to improve the accuracy of the Bayesian classifier model, command detectors are also used to extract further features that are useful identifiers of potentially malicious activities. Command detectors also inspect resources, assets and libraries as these can be used to hosts malicious scripts or hidden payloads that may not be detected by scrutinizing the application code logic alone. Note that these, unlike API calls, can be located within scripts of raw resources or assets or even within external libraries. Attributes searched for by the command detectors include system commands like 'chmod', 'mount', 'remount' 'chown'. Others include the Java Realtime.exec command through which persistent background child processes containing malicious payload can be launched. The command detector also matches Android commands like 'pm install', which can enable stealthy installation of additional packages.

*3) Permissions detectors*

These are used to glean information about the permissions requested by the application declared within the manifest file. In addition to providing information that could build features for the Bayesian classification, the permission detectors also provide information-rich profiling of the applications, which can give insight for further analyses if needed.

## IV. BAYESIAN CLASSIFICATION MODEL

Data mining technologies are becoming increasingly important in the anti-malware industry, particularly in augmenting well-established *heuristics* and *generics* methods [24]. Data mining drives automation, which is motivated by reducing maintenance costs associated with the traditional heuristics and generics methods [24]. Data mining employs machine learning methods for inference, prediction, classification etc. Hence, it is important to select an appropriate method depending on the particular application. Bayesian classification is well suited to our problem of filtering large amounts of apps as it can perform relatively fast classification with low computational overhead once trained. Another important property which motivates its implementation in our approach for detecting suspicious Android applications, is the ability to model both an 'expert' and 'learning' system with relative ease compared to other machine learning techniques. Bayesian method allows the incorporation of prior probabilities (expert knowledge) even before the training phase. This hybrid property can be exploited to optimize the classifier's performance without incurring additional computational overhead. However, in this research, only its applicability as a pure 'learning' method is explored. Evaluation of the hybrid property is outside the scope of this paper.

### A. The classifier model

The Bayesian based classifier consists of learning and detection stages. The learning stage uses a training set in form of known malicious samples in the wild and benign Android applications, collectively called the app corpus. The Java-based package analyzer uses the detectors to extract the desired features from each app in the corpus. The feature set is subsequently reduced by a feature reduction function while the training function calculates the likely probability of each selected feature occurring in the malicious and benign applications. The training function also calculates the prior probability of each class i.e. *suspicious* and *benign*.

The feature sets along with their likelihood probabilities are stored for use in the subsequent classification stage. New applications are assessed in the classification stage with respect to the features in the selected feature set.

### B. Feature ranking and selection

Let an application characteristic $r_i$ obtained from the APK analyzer detectors be defined by a random variable:

$$R_i = \begin{cases} 1, & \text{if discovered by the detectors} \\ 0, & \text{otherwise} \end{cases} \quad (1)$$

In order to ensure selection of the most relevant application features, Mutual Information (MI) [25] calculation is utilized to rank the extracted features within the feature reduction function. Let C be a random variable representing the application class, *suspicious* or *benign*:

$$C \in \{suspicious, benign\}$$

Every application is assigned a vector defined by $\vec{r} = (r_1, r_2, ... r_n)$ with $r_i$ being the result of the $i$-th random variable $R_i$.

As the goal is to select the most relevant features, the feature reduction function calculates the Mutual Information of each random variable, thus:

$$MI(R_i, C) = \sum_{r \in \{0,1\}} \sum_{c \in \{sus, ben\}} P(R_i=r; C=c) \cdot \frac{P(R_i=r; C=c)}{P(R_i=r).P(C=c)} \quad (2)$$

After calculating the MI for each feature, the feature set is then ranked from largest to smallest in order to select those that maximize the MI between $R$ and $C$ thus enabling optimum classifier performance.

### C. Bayesian classification

According to Bayes theorem, the probability of an application with the feature vector $\vec{r} = (r_1, r_2, ... r_n)$ belonging in class C is defined by:

$$P(C = c \mid \vec{R} = \vec{r}) = \frac{P(C = c) \prod_{i=1}^{n} P(R_i = r_i \mid C = c)}{\sum_{j \in \{0,1\}} P(C = c_j) \prod_{i=1}^{n} P(R_i = r_i \mid C = c_j)} \quad (3)$$

Where $P(R_i = r_i \mid C = c)$ and $P(C = c_j)$ are the estimated frequencies calculated on the app learning corpus. While $n$ is the number of features used in the classification engine; $c_0$ and $c_1$ are the benign and suspicious classes respectively.

An app represented by the vector $\vec{r} = (r_1, r_2, ... r_n)$ is classified as benign if:

$$P(C = benign \mid \vec{R} = \vec{r}) > P(C = suspicious \mid \vec{R} = \vec{r}) \quad (4)$$

Otherwise, it is classified as suspicious. In terms of classification error, two cases can occur: (a) A benign app misclassified as suspicious. (b) A suspicious app misclassified as benign. In the context of our problem, the latter case is more critical since allowing a malicious app to reach an end device is more critical than excluding a benign app from the distribution chain to be subject to further scrutiny.

### D. Evaluation measures

To evaluate the predictive accuracy of classifiers, several measures have been proposed in the literature. In the context of our problem the relevant measures utilized in our experiments are given below.

Let $n_{ben \to ben}$ be the number of benign applications correctly classified as benign, $n_{ben \to sus}$ the number of misclassified benign applications, $n_{sus \to sus}$ the number of suspicious applications correctly identified as suspicious while $n_{sus \to ben}$ represents the number of misclassified suspicious applications. Accuracy and Error Rate are respectively given by:

$$Acc = \frac{n_{ben \to ben} + n_{sus \to sus}}{n_{ben \to ben} + n_{ben \to sus} + n_{sus \to ben} + n_{sus \to sus}} \quad (5)$$

$$Err = \frac{n_{ben \to sus} + n_{sus \to ben}}{n_{ben \to ben} + n_{ben \to sus} + n_{sus \to ben} + n_{sus \to sus}} \quad (6)$$

We also define *the false positive rate (FPR), false negative rate (FNR), true positive rate (TPR), true negative rate (TNR)* and *precision (p)* as follows:

$$FPR = \frac{n_{ben \to sus}}{n_{ben \to sus} + n_{ben \to ben}} \quad (7)$$

$$FNR = \frac{n_{sus \to ben}}{n_{sus \to sus} + n_{sus \to ben}} \quad (8)$$

$$TPR = \frac{n_{sus \to sus}}{n_{sus \to ben} + n_{sus \to sus}} \quad (9)$$

$$TNR = \frac{n_{ben \to ben}}{n_{ben \to sus} + n_{ben \to ben}} \quad (10)$$

$$p = \frac{n_{sus \to sus}}{n_{ben \to sus} + n_{sus \to sus}} \quad (11)$$

## V. METHODOLOGY AND EXPERIMENTS

As mentioned earlier, our implementation of an APK analyzer includes the steps illustrated in Figure 2. A total of 2000 applications were analyzed in order to extract the features which are then used by the feature selection function to provide a relevance ranking according to the MI equation (2) given earlier. The 2000 APKs consisted of 1000 malware samples from 49 different families and 1000 benign apps downloaded from official and third party Android markets.

The breakdown of the 49 malware families used and their respective number of samples are shown in Table I. The malware samples were obtained from the Android Malware Genome Project [3].

The set of 1000 non-malicious apps were made up of different categories in order to cover a wide variety of application types. The categories include: entertainment, tools, sports, health and fitness, news and magazines, finance, music and audio, business, education, games and a few other miscellaneous categories. The apps from third party market places were screened using *virustotal* scanning service and by manual inspection of extensive profile sets created by our analyzer so as to discard any apps of highly suspicious nature from the set.

A total of 58 feature attributes were defined as matching criteria for the property detectors. 10 out of these did not yield any match in both benign and malware sets and so were discarded. The remaining 48 were subsequently applied to the feature selection function which ranked them according to the calculated MI. The top 25 features and their respective frequencies in both sets are shown in Table II.

TABLE I. MALWARE FAMILIES USED AND THEIR NUMBERS

| Family | No of samples | Family | No of samples |
|---|---|---|---|
| ADRD | 22 | GingerMaster | 4 |
| AnserverBot | 130 | GoldDream | 47 |
| Asroot | 8 | Gone60 | 9 |
| BaseBridge | 100 | GPSSMSSpy | 6 |
| BeanBot | 8 | HippoSMS | 4 |
| Bgserve | 9 | Jifake | 1 |
| Coinpirate | 1 | jSMSHider | 16 |
| CruseWin | 2 | KMin | 52 |
| DogWars | 1 | LoveTrap | 1 |
| DroidCoupon | 1 | NickyBot | 1 |
| DroidDeluxe | 1 | NickySpy | 2 |
| DroidDream | 16 | Pjapps | 58 |
| DroidDreamLight | 46 | Plankton | 11 |
| DroidKungFu1 | 30 | RougeLemon | 2 |
| DroidKungFu2 | 34 | RougeSPPush | 9 |
| DroidKungFu3 | 144 | SMSReplicator | 1 |
| DroidKungFu4 | 80 | SndApps | 10 |
| DroidKungFuSapp | 3 | Spitmo | 1 |
| DroidKungFuUpdate | 1 | Tapsnake | 2 |
| Endofday | 1 | Walkinwat | 1 |
| FakeNetflix | 1 | YZHC | 22 |
| FakePlayer | 6 | zHash | 11 |
| GamblerSMS | 1 | Zitmo | 1 |
| Geinimi | 69 | Zsone | 12 |
| GGTracker | 1 | | |

TABLE II. TOP SELECTED FEATURES AND THEIR FREQUENCIES IN BENIGN AND MALWARE SETS OF 1000 SAMPLES IN EACH CATEGORY

| Features | Benign | malware |
|---|---|---|
| getSubscriberId (TelephonyManager) | 42 | 742 |
| getDeviceId (TelephonyManager) | 316 | 854 |
| getSimSerialNumber (TelephonyManager) | 35 | 455 |
| .apk (secondary payload) | 89 | 537 |
| intent.action.BOOT_COMPLETED | 69 | 482 |
| chmod (system command) | 19 | 389 |
| Runtime.exec( ) (Executing process) | 62 | 458 |
| abortBroadcast (intercepting broadcast notifications) | 4 | 328 |
| getLine1Number (TelephonyManager) | 111 | 491 |
| /system/app | 4 | 292 |
| /system/bin | 45 | 368 |
| createSubprocess (creating child process) | 0 | 169 |
| getSimOperator (TelephonyManager) | 37 | 196 |
| remount (system command) | 3 | 122 |
| DexClassLoader (stealthily loading a class) | 16 | 152 |
| pm install (installing additional packages) | 0 | 98 |
| getCallState (TelephonyManager) | 10 | 119 |
| chown (system command) | 5 | 107 |
| .jar (secondary payload) | 87 | 252 |
| mount (system command) | 29 | 152 |
| KeySpec (code encryption) | 99 | 254 |
| /system/bin/sh | 4 | 90 |
| SMSReceiver | 3 | 66 |
| getNetworkOperator (TelephonyManager) | 202 | 353 |
| SecretKey (code encryption) | 119 | 248 |

Table II shows that some of the properties were only found to be present in the malware samples, while references to system commands mainly appeared in the malware class. Although the Telephony manager API calls are present in both classes as expected, higher occurrences were observed in malware samples. The references to .apk and .jar files, which are meant to detect the presence of secondary apps are found in both classes with more in the malware set. Whilst secondary apps can be used to hide malicious payload, some legitimate apps such as popular ad and mobile payment frameworks also utilize them. Notwithstanding the use of some properties by both malicious and legitimate apps, Bayesian classification provides the ability to probabilistically combine several of them including those that are almost exclusively found in one class only to enable more effective discrimination.

### A. Bayesian Classifier training

For the training of the classification model, the same set of 2000 samples comprising 1000 malware and 1000 benign apps were used. In order to provide for testing and evaluation according to the earlier defined evaluation criteria in equations (5) to (11), 5-fold cross validation is employed. Thus, 1600 samples (800 each of benign and malware) were used in the training, while the remaining 400 (200 each of benign and malware) were used for testing. Hence, the experiments undertaken used 5 different training and testing sets each containing a different testing portion with samples outside of its own training portion. This strategy was chosen to provide a wider range of samples for the testing of the classifier.

## VI. RESULTS AND DISCUSSIONS

Figures 3 to 6 depict the results of the experiments undertaken to evaluate the Bayesian classifier. Five different sets of features were used containing 5, 10, 15 and 20 features respectively, to obtain the results depicted in Figs. 3 to 5 and Table III. 10f, 15f and 20f refer to the top 10, top 15 and top 20 ranked features respectively. 5fT refers to five top features while 5fL refers to the five lowest ranked features from the set of the selected 20 (i.e. 16th to 20th ranked).

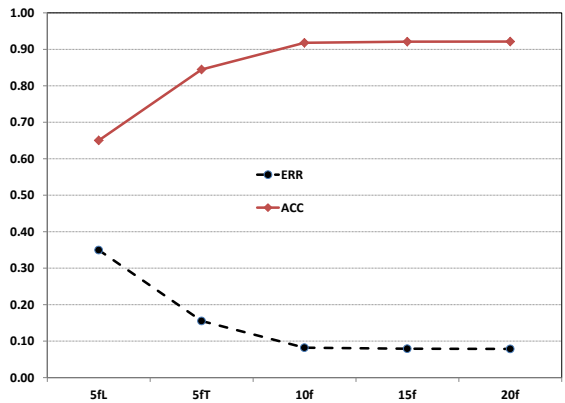

Figure 3. Average error rates and accuracy values for different feature sets.

From Figure 3, we observe increasing accuracy and decreasing error rates when a larger number of features are used to train the classifier. A very close performance between 15f and 20f can be seen, which indicates that only marginal improvement would accrue from employing a larger number of features for training the model. An interesting observation is the difference in 5fT and 5fL performance, with 5fT achieving close to 85% average accuracy compared to 65% with 5fL. This highlights the effectiveness of the ranking by the feature selection function of the analyzer, since the same number of features but of different rankings were present in 5fL and 5fT.

Figure 4 depicts the true negative rate and the false positive rates observed in the experiments. The results do not follow the same trend observed with accuracy and error rate. Instead, 5fL achieves the highest TNR of 0.954. The reason for this is that when trained with a relatively small number of features, there will be more occurences of feature vectors $\vec{r} = (r_1, r_2, ... r_n)$ with no detected features, i.e. zero vectors $\vec{r} = (0, 0, ... 0)$. And since a larger number of these vectors will come from benign apps, this has the effect of strengthening the TNR. But in accordance with equations (5) and (6), overall accuracy falls while error rate increases because the false negatives $n_{sus \rightarrow ben}$ also goes up as a result. This is because the classifier will have prior knowledge of $\vec{r} = (0, 0, ... 0)$ with higher probability of falling within 'benign' than 'malicious'.

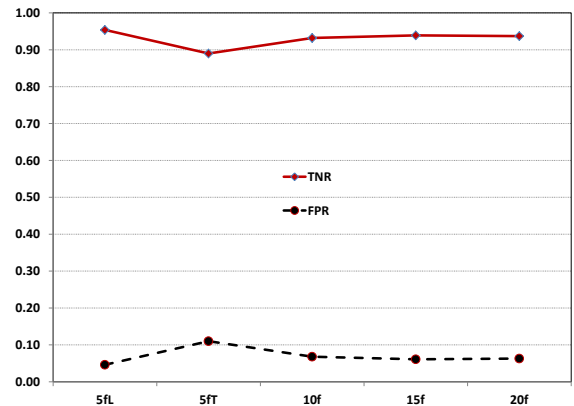

Figure 4. Average true negative rates and false positive rates for the different feature sets.

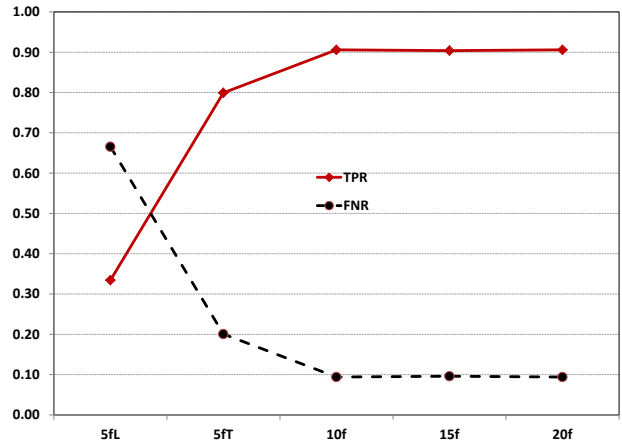

Figure 5. Average true positive rates and false negative rates for the various feature sets.

TABLE III.     SUMMARY OF EXPERIMENTAL RESULTS FROM THE
CLASSIFICATION MODEL FOR DIFFERENT FEATURE SETS.

|     | ERR   | ACC   | TNR   | FPR   | TPR   | FNR   | Prec. | AUC     |
|-----|-------|-------|-------|-------|-------|-------|-------|---------|
| 5fL | 0.350 | 0.650 | 0.954 | 0.046 | 0.335 | 0.665 | 0.860 | 0.61709 |
| 5fT | 0.155 | 0.845 | 0.890 | 0.110 | 0.799 | 0.201 | 0.880 | 0.94437 |
| 10f | 0.082 | 0.918 | 0.932 | 0.068 | 0.906 | 0.094 | 0.931 | 0.97428 |
| 15f | 0.079 | 0.921 | 0.939 | 0.061 | 0.904 | 0.096 | 0.937 | 0.97232 |
| 20f | 0.079 | 0.921 | 0.937 | 0.063 | 0.906 | 0.094 | 0.935 | 0.97223 |

Again, from Figure 4, there is not much to choose between 15f and 20f for TNR and FPR results. Although the false positive rates observed were relatively low (around 6%), FPR is not considered as critical as FNR since the latter directly affects the proportion of malware that will be 'missed'. On the other hand, a low FPR means that less benign apps will need to be subject to further scrutiny; and it then becomes more cost-effective or less time consuming to do so when the FPR is low, given a massive amount of apps.

From Figure 5, it can be seen that 10f, 15f and 20f sets yielded TPRs of 0.906, 0.904 and 0.906 respectively. Correspondingly their FNR values are 0.094, 0.096 and 0.094. It is worth emphasizing that these TPRs indicate a significantly higher malware detection rate than were achieved by signature-based anti-virus scanners as reported in [3], which were tested on the same malware sample set. The study reported a best case of 0.796 detection rate while 0.202 was the worst case amongst four signature-based mobile AVs. The paper stated that most of the undetected malware were unknown samples whose signatures were unavailable at the time of study. Nevertheless, the Bayesian learning approach presented in this paper has demonstrated capability of detecting unknown malware since the training and testing sets are of unique samples.

Because FNR is critical in the context of our problem, the results warrant further discussion. We observed that some of false negatives that occurred resulted from 'zero feature vectors'. While increasing the number of features from 20 could reduce the zero feature vector occurrences, this will not necessarily result in better classification accuracy (as Figure 3 illustrates). One way to overcome this could be through combining related features as matching criteria, for example the .apk and .jar (secondary payload matching features). This will allow new features to be introduced into the classification model while keeping an optimum total required to train the model.

We also note from Table III that precision and AUC also mirror the TPR and accuracy results by generally improving with higher number of features, with those of 10f, 15f, and 20f quite close to one another. Precision (expressed in equation 11), reflects the precision of the model when classifying samples as suspicious. AUC (Area Under Curve), on the other hand, is the total area under the Receiver Operation Characteristic (ROC) curve which is a plot of TPR vs. FPR for every possible detection cut-off. An AUC of 1 implies perfect classification. Therefore, an AUC value closer to 1 denotes better classifier predictive power.

In Figure 6 and Table IV, results obtained from investigating the effect of varying numbers of training samples on the performance metrics are presented. These are from the 20 top feature set with 100, 250, 500, 1000 and 1600 training samples evaluated over the same testing sets using 5-fold cross validation. Each training set had 50% benign and 50% malware. Note from Figure 6 that while increasing the number of training samples had a noticeable effect on the TPR, it had much less impact on the TNR. Increasing the number of training samples improved the TPR and the accuracy also got better as a result. The reason for this observation can be explained by the fact that there is greater variability in the characteristics of the malware samples with respect to the 20 selected features compared to the benign samples. Hence, varying the number of training samples has more marked effect on the metrics linked to the malware class (i.e. TPR and FNR) than those related to the benign class (i.e. TNR and FPR). This can be seen more clearly from Table IV. We can therefore conclude that a larger training sample set, especially with more malware samples, will further improve the TPR, lower the FNR and in turn improve the overall accuracy of the classifier.

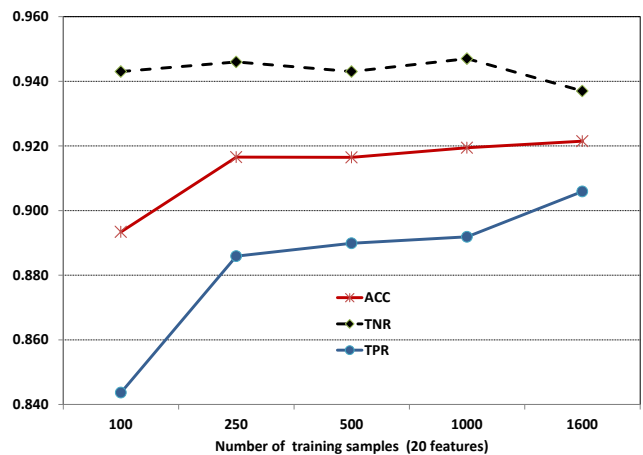

Figure 6.  Average true negative rates, accuracy, and true positive rates for different training samples with 20 features.

TABLE IV.     SUMMARY OF RESULTS FOR DIFFERENT TRAINING SAMPLES
WITH 20 FEATURES.

| Samples | ERR   | ACC   | TNR   | FPR   | TPR   | FNR   | Prec. | AUC     |
|---------|-------|-------|-------|-------|-------|-------|-------|---------|
| 100     | 0.107 | 0.893 | 0.943 | 0.057 | 0.844 | 0.156 | 0.937 | 0.95794 |
| 250     | 0.083 | 0.917 | 0.946 | 0.054 | 0.886 | 0.114 | 0.943 | 0.96877 |
| 500     | 0.084 | 0.916 | 0.943 | 0.057 | 0.890 | 0.110 | 0.940 | 0.97119 |
| 1000    | 0.081 | 0.919 | 0.947 | 0.053 | 0.892 | 0.108 | 0.944 | 0.97177 |
| 1600    | 0.079 | 0.921 | 0.937 | 0.063 | 0.906 | 0.094 | 0.935 | 0.97223 |

Precision is affected by both false positives and true positives (see equation 11). Hence, the results are variable with respect to changing number of training samples, as can be seen from Table IV. AUC (i.e. classifier predictive power) on the other hand, improves with the increase in number of training samples.

## VII. Conclusion

In this paper we proposed and evaluated a machine learning-based approach for detecting Android malware. In particular, a novel application of Bayesian classification is applied to this problem. Through reverse engineering of the Android applications using an APK analyzer implemented in Java, a set of 58 properties were extracted to provide features that are subsequently ranked by a feature selection function. The properties are extracted from the applications by means of detectors that search for patterns and references to API calls, system commands, etc. frequently encountered with Android malware.

1000 samples from 49 Android malware families together with another 1000 benign applications across a variety of categories have been used for feature extraction and training of the Bayesian classifier. From the experiments conducted, it is discovered that 15 to 20 features are sufficient to provide optimum performance based on the detected properties upon which the features are based and the ranking by the feature selection function.

The results presented in the paper showed significantly better detection rates than were achieved by popular signature-based antivirus software tested previously on the same set of malware samples used in our experiments. The malware samples used in the experiments were from the largest publicly available collection at the time of writing. Future work could investigate the classifier performance with larger sample sets as more malware samples are discovered in the wild. Further studies could also investigate performance improvement via prior incorporation of expert knowledge within the Bayesian classification model.